\documentstyle[preprint,aps,epsf]{revtex}
\title{Quantum Stability of (2+1)-Spacetimes with Non-Trivial Topology}
\author{Masaru Siino\footnote{e-mail: 
msiino@tap.scphys.kyoto-u.ac.jp, 
JSPS fellow}\\
\it Department of Physics, Kyoto University\\
Kitashirakawa, Sakyoku, Kyoto 606-01, Japan}

\begin{document}
\maketitle
\begin{abstract}
Quantum fields are investigated in the (2+1)-open-universes with non-trivial 
topologies by the method of images. The universes are locally de Sitter spacetime and
anti-de Sitter spacetime. In the present article we study spacetimes whose spatial 
topologies are a torus with a cusp and a sphere with three cusps as a step toward the 
more general case. A quantum energy momentum tensor is obtained by the point stripping method. 
Though the cusps are no singularities, the latter cusps cause the divergence of the quantum field.
This suggests that only the latter cusps are quantum mechanically unstable. Of course at 
the singularity of the background spacetime the quantum field diverges. 
Also the possibility of the divergence of topological effect by a negative spatial curvature 
is discussed.  Since the volume of the negatively curved space is larger than that of 
the flat space, one see so many images of a single source by the non-trivial topology. 
It is confirmed that this divergence does not appear in our models of topologies. 
The results will be applicable to the case of three dimensional multi black hole\cite{BR}.
\end{abstract}

\section{Introduction}
When we consider matter fields in a spacetime with a non-trivial topology, the boundary effects of quantum fields 
appear. This is one of main targets in quantum field theory in 
curved spacetimes. Such effects have been studied well in spatially flat 
spacetimes\cite{DOW}, but not so well in spatially curved spacetimes. 
This is because of the 
complexity of the topology which is allowed in such curved spatial 
section. In other words, there 
will be new topological effects of quantum field in the curved spatial section with a little complex topology.

To construct a space with a non-trivial topology, we identify the  
points of covering space by the discrete subgroup of the isometry of the space. Then we 
want to consider the  covering space with an appropriately simple 
isometry group such that the topology has some interesting characteristics. As the 
space with simple isometry, there are $S^n$, $R^n$ and $H^n$, so called closed-, flat- and 
open-universe. Furthermore, we decide to treat $H^n$ since this hyperbolic 
space allows various topologies possessing interesting characteristics.

To treat this open-universe as a background spacetime, we determine its time evolution. For 
simplicity we consider maximally symmetric spacetimes of de Sitter spacetime 
with a hyperbolic chart or anti de Sitter spacetime in a Robertson-Walker 
coordinate. Their spatial sections are $H^n$. The de Sitter spacetime 
with the hyperbolic chart may be 
important in a cosmological sense. It is believed that the global 
feature of our universe is homogeneous and isotropic. If our observation 
suggests that the spatial curvature of our universe is negative, the 
background spacetime is locally open-universe. In the inflation, the de 
Sitter spacetime with a hyperbolic chart is a good model for 
cosmology\cite{TN}. If we prefer a 
universe with a finite volume, the de Sitter spacetime with a hyperbolic 
chart with non-trivial topology will become important.

The topology of the open-universe (in the present article, the 
`open-universe' means not an open topology but only a negatively curved 
space) is well known in a two 
dimensional space. Then we construct the simple example of a two dimensional 
open-universe with interesting topology in (2+1)-de Sitter spacetimes and (2+1)-anti de Sitter spacetimes.  
Quantum scalar field is studied in these spacetimes using point 
stripping manner.
The divergences of the quantum fields will be discussed.

In the section 2, we prepare simple model of the universes with 
interesting topologies in the (2+1)-de Sitter spacetime and the (2+1)-anti de 
Sitter spacetime. The quantum field is investigated in the section 3. The 
last section is devoted to a summary and discussions. 
\section{(2+1) Open-universe with Non-trivial Topology}
\subsection{Two Dimensional Universe}
\begin{figure}[tbp]
	\centerline{\epsfbox{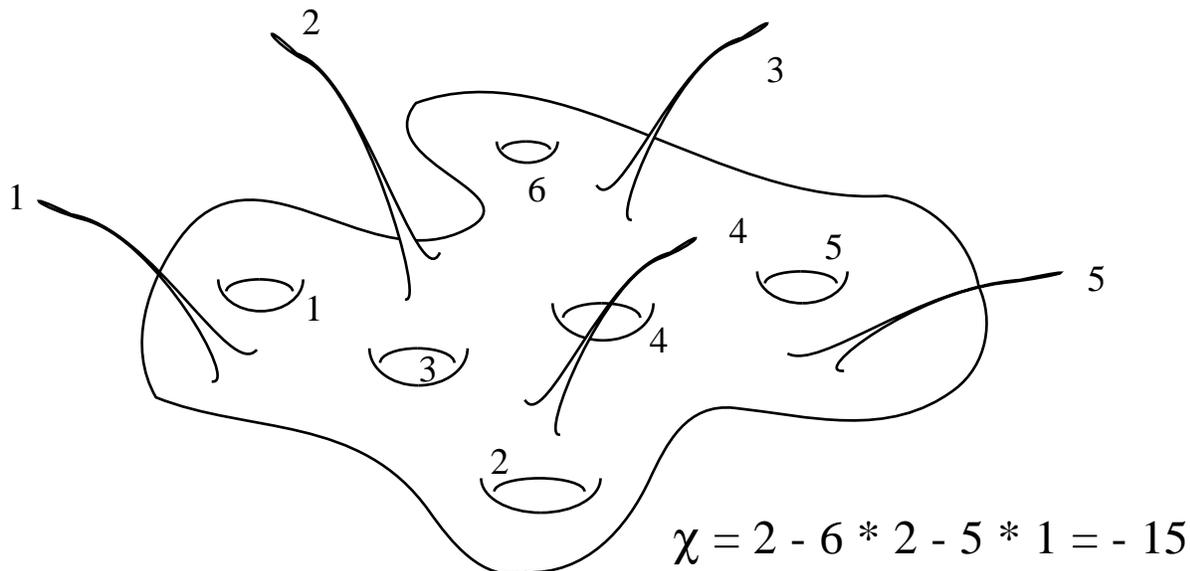}}
	\caption{A example of a two dimensional sphere with a finite volume is shown. There are six handles and five cusps. Its Euler number is $\chi = 2-2*6-1*5=15$.}
	\protect\label{fig:ex}
\end{figure}

First of all, we develop topologies of two-dimensional spatial sections.
For the simplicity of topologies we treat (2+1)-spacetime in the present 
article. 
For a cosmological reason and a further simplicity, spatial two-dimensional sections are
assumed to be $S^2$, $R^2$ or $H^2$ corresponding to closed-, flat- or 
open-universe, respectively. In a two-dimensional space, the topologies of 
complete manifolds are classified by Euler numbers. This is calculated from 
the number of handles and cusps (see Fig.\ref{fig:ex}). The cusp is a 
point at infinity with a needle-like structure. Here it should be 
emphasized that the cusp is no unnatural or artificial. These points at 
infinity are no singularity. There is no reason to give them special 
treatment in a classical physics\cite{MS}. 

From 
the Gauss-Bonnet theorem for a complete 2-manifold, the Euler number $\chi$ 
is given by
\begin{equation}
	{1\over 4\pi}\int dv \thinspace^{(2)}R = \chi =2-2 N(handle)-N(cusp),
	\label{}
\end{equation}
where $\thinspace^{(2)}R$ is a two-dimensional scalar curvature and $N(\ast)$ is 
the number of $\ast$.
The signature of $\thinspace^{(2)}R$ restricts the variety 
of the topology. Since the Euler number is less than $2$ ($\chi =2$ is 
for a sphere), the case of negative curvature allows various topologies 
(various numbers of handles and cusps). In such negative  
curvature space, we expect new topological effects of a quantum field. 
Since the cusp is an infinitely small structure, it may cause the 
divergence of the quantum field.  
The negative curvature space has a crucial characteristic that 
the volume of the space is larger than that of the flat space at a distant region.
Then one will see so many images of sources because of a non-trivial 
topology. The method of images may suffer the difficulty of a 
divergence. To discuss these speculations, we must treat general 
topologies of the negative curvature space $H^2$. In the present article, 
however, only two simple cases of them can be investigated since these 
cases possess the cusps and the above mentioned characteristic.

To construct a non-trivial topology of the negative curvature space, we draw a polygon surrounded by 
geodesics on a hyperbolic space $H^2$ as a fundamental region and identify the 
geodesics. Poincar\'{e} model is one of the models of the hyperbolic 
space, which is conformally flat and a compact chart. The metric of the 
Poincar\'{e} model is 
\begin{equation}
 	ds^2={4 (dr^2 + r^2 d\phi^2) \over (1-r^2)^2},
 	\label{eqn:pm}
\end{equation} 
whose spatial curvature is $-1$. In this model, geodesics are circles 
crossing a circle with $r=1$ at right angles. This circle with $r=1$ is a sphere at 
infinity corresponding to the infinity of $H^2$. 

\begin{figure}[tbp]
	\centerline{\epsfbox{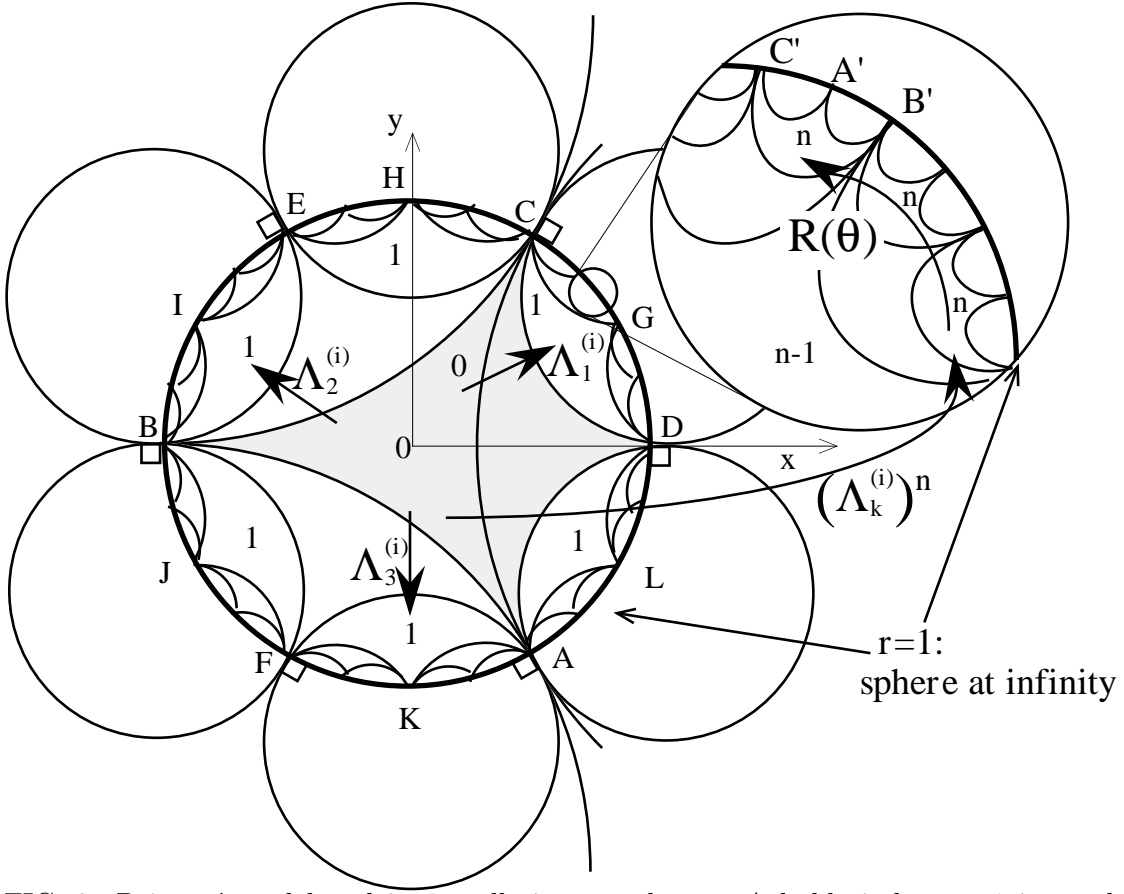}}
	\caption{Poincar\'{e} model and its tessellation are shown. A bold circle $r=1$ is a sphere at infinity. A shaded region is a fundamental region of a torus with a cusp or a sphere with three cosps. $\triangle ABC$ is transformed to a $n$-th outward position by $(\Lambda^{(i)}_k)^n$. Furthermore, $R(\theta)$ transform it  to $\triangle A'B'C'$.}
	\protect\label{fig:ttr}
\end{figure}

\begin{figure}[tbp]
	\centerline{\epsfbox{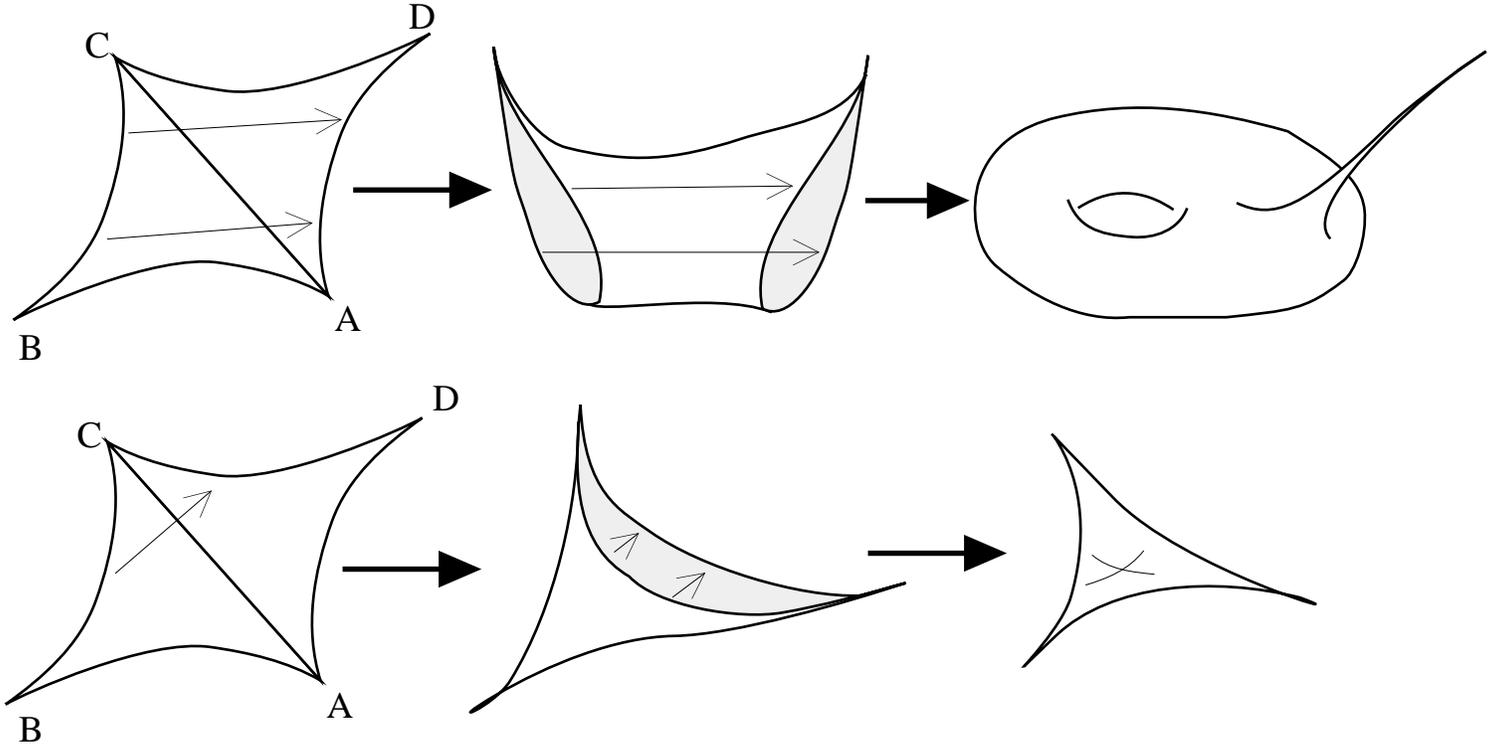}}
	\caption{The upper shows that a torus with a cusp is constructed from a tetragon. The lower shows a tetragon becomes a sphere with three cusps.}
	\protect\label{fig:con}
\end{figure}

Now we give simple topologies including cusps. The simplest polygon 
producing non-trivial topology is a tetragon.
Drawing the tetragon $\Box ABCD$ of a fundamental region as in Fig.\ref{fig:ttr}, the tessellation of $H^2$ 
by this tetragon becomes simplest. In Fig.\ref{fig:ttr}, a half 
of the tetragon, $\triangle ABC $ and $\triangle ACD$, is a regular 
triangle and its vertices $A,B,C,D$ are on the sphere at 
infinity.  $H^2$ are 
tessellated by parallel transformations of these triangles. The tessellation turns out to be a series of regular polygons 
whose centers are the origin of the Poincar\'{e} model. 
The triangles form a self-homothetic structure.
Requiring orientability, there are two pairs of identifications for 
geodesics contained by the triangles, which provide complete manifold.
One is 
\begin{eqnarray}
	\triangle ABC &\longrightarrow & \triangle DCG \ \ \ {\rm by} \ 
	\Lambda^{(1)}_{DCG}\\
	\triangle ABC &\longrightarrow & \triangle LAD \ \ \ {\rm by} \ 
	\Lambda^{(1)}_{LAD},\ \ \ \Lambda^{(1)}\in SO(2,1).
	\label{}
\end{eqnarray}
They generate a discrete subgroup $\Gamma^{(1)}$ of $ SO(2,1)$. By these 
identifications, the topology of $H^2/\Gamma^{(1)}$ becomes a torus having a point at 
infinity, which is a cusp (see Fig.(\ref{fig:con})).
 Of course, the Gauss-Bonnet theorem gives its Euler number as
\begin{equation}
	\chi = 2-2 N(handle)-N(cusp)=2-2-1=-1={1\over 4\pi}\int_{H^2/\Gamma^{(1)}} 
	dv\thinspace^{(2)}R.
	\label{}
\end{equation}
The other is 
\begin{eqnarray}
	\triangle ABC &\longrightarrow & \triangle GDC \ \ \ {\rm by}\  \Lambda^{(2)}_{GDC}\\
	\triangle ABC &\longrightarrow & \triangle ADL \ \ \ {\rm by}\  
	\Lambda^{(2)}_{ADL},\ \ \ \Lambda^{(2)}\in SO(2,1).
	\label{}
\end{eqnarray}
They generate $\Gamma^{(2)}\subset SO(2,1)$. A resultant topology 
$H^2/\Gamma^{(2)}$ is a 
sphere with three cusps (Fig.\ref{fig:con}). The Euler number of it is
\begin{equation}
	\chi = 2-2 N(handle)-N(cusp)=2-0-3=-1={1\over 4\pi}\int_{H^2/\Gamma^{(2)}} dv\thinspace^{(2)}R.
	\label{}
\end{equation}
It is noted that these manifolds $H^2/\Gamma^{(1)}$ and 
$H^2/\Gamma^{(2)}$ are easier to handle than the double-torus which is a well known example of 
hyperbolic manifolds. For instance, 
though the fundamental group of our manifolds is $(a,b)$, 
that of the double torus is $(a,b,c,d:aba^{-1}b^{-1}cdc^{-1}d^{-1}=1)$\cite{NS}.

Here we express all elements of $\Gamma^{(i)}$ in an appropriate form 
for the rest of the present article. Using $\{\Lambda^{(i)}_1, \Lambda^{(i)}_2, 
\Lambda^{(i)}_3\}$ acting as
\begin{eqnarray}
	 \triangle ABC& \longrightarrow & \triangle DCG\ \ \  {\rm by} \ 
	 \Lambda^{(1)}_1
	\label{} \\
	  & \longrightarrow & \triangle BIE\ \ \  {\rm by} \ 
	 \Lambda^{(1)}_2
	\label{} \\
	  & \longrightarrow & \triangle KFA\ \ \  {\rm by} \  
	 \Lambda^{(1)}_3
	\label{} \\
	  & \longrightarrow & \triangle GDC\ \ \  {\rm by} \ 
	 \Lambda^{(2)}_1
	\label{} \\
	  & \longrightarrow & \triangle EBI\ \ \  {\rm by} \ 
	 \Lambda^{(2)}_2
	\label{} \\
	  & \longrightarrow & \triangle AKF\ \ \  {\rm by} \ 
	 \Lambda^{(2)}_3,
	\label{}
\end{eqnarray}
 $\thinspace^\forall T^{(i)}\in\Gamma^{(i)}$ is given by
\begin{equation}
	T^{(i)}=R(\theta_j)(\Lambda^{(i)}_k)^n.
	\label{eqn:dt}
\end{equation}
For example, $T^{(i)}$ transform $\triangle ABC$ to $\triangle A'B'C'$ in 
Fig.\ref{fig:ttr}. $\triangle ABC$ is transformed to a $2n$-th outward position 
of triangles by $(\Lambda^{(i)}_k)^n$
and rotated around the origin by $R(\theta_j)$ with an appropriate angle $\theta_j$. 
$k$ is selected from $1\sim 3$ so that the orders of the vertices 
match between $\triangle ABC$ and $\triangle A'B'C'$.

 For the rest of the article, we give $\Lambda^{(1)}_1$ in terms of 
the Poincar\'{e} model. Using $z=r e^{i\theta}$, this representation of 
the isometry of $H^2$ becomes 
a subgroup of $SL(2,C)$ in the coordinate of the Poincar\'{e} model and is given by
\begin{equation}
	z\rightarrow f(z)={a z +b\over \bar{b}z +\bar{a}},\ \ \  
	\left(\begin{array}{cc}
		 a & b  \\		
		\bar{b} & \bar{a}
	\end{array}\right)\in SL(2,C).
	\label{eqn:sl}
\end{equation}
$\Lambda^{(1)}_1$ is given by
\begin{eqnarray}
	f(z)&=&{a_0 z +b_0\over \bar{b_0}z +\bar{a_0}}\label{eqn:n1}\\
	a_0&=&{1\over\sqrt{1-r_0^2}}\sqrt{i{r_0 e^{i\pi/8}-1 \over r_0 
	e^{-i\pi/8}-1}},\ \ \  b_0={1\over\sqrt{1-r_0^2}}{r_0 
	e^{i\pi/8}\over a_0}, \label{eqn:n2}\\
	r_0&=&{1-\sin{\pi \over 8} \over \cos{\pi \over 8}}
	\label{eqn:n3}
\end{eqnarray}
Here we should note that $\Lambda^{(1)}$'s and $\Lambda^{(2)}$'s are in the different 
category of the Lorentz group $SO(2,1)$. It is known that all elements of 
$SO(2,1)$ are $SO(2,1)$-conjugate to an element of the 
following forms. We call them as standard forms.
In the $SL(2,C)$ representation (\ref{eqn:sl}),
\begin{itemize}
	\item  1) An elliptic element is conjugate to
	\begin{equation}
	T_e:\ \ f(z)=e^{i\theta}z={e^{i\theta/2}z\over e^{-i\theta/2}}
	\end{equation}
	with one fixed point on $H^2$.
	\item  2) A parabolic element is conjugate to
	\begin{equation}
	T_p:\ \ f(z)={(1+i)z + i \over -i z +(1-i)}
	\label{eqn:t2}
	\end{equation}
	with one fixed point on the sphere at infinity.	
	\item  3) A hyperbolic element is conjugate to
	\begin{equation}
	T_h:\ \ f(z)={z\cosh\beta  +\sinh\beta \over z\sinh\beta  + \cosh\beta },
	\label{eqn:tx}
	\end{equation}
	with two fixed point on the sphere at infinity.
\end{itemize}
These angle parameters $\theta, \beta$ are real numbers.
Though $\Lambda^{(1)}$'s are conjugate the category $3)$, $\Lambda^{(2)}$'s are 
parabolic. We note that the fixed point of a parabolic element 
corresponds to a cusp produced by the parabolic element. It is revealed 
in the next section that these facts affect the quantum field.

\subsection{de Sitter and anti de Sitter Spacetime with Non-trivial Topology}
Now we consider (2+1)-spacetime whose certain spatial sections have above mentioned 
topologies. For simplicity, Teichm\"{u}ller deformation\cite{FS} is not 
considered and every 
identifications are supposed to be done on the spatial 2-sections of the synchronous gauge. 
Then the local 
geometry of the spacetime is that of the open FRW-universe.
For further simplicity, we assume the spacetime is maximally symmetric. 
Such (2+1)-spacetimes allowing spatial $H^2$-sections are Minkowski 
spacetime, de Sitter 
spacetime $dS^3$ and anti de Sitter spacetime $AdS^3$. Though we treat 
only $dS^3$ and $AdS^3$ in the present article, the investigation of the
present article is easily applicable to the case of the Minkowski spacetime.

The isometry groups of $dS^3$ and $AdS^3$ are $SO(3,1)$ and $SO(2,2)$, 
respectively. The open chart of $dS^3$ and the 
RW(Robertson-Walker)-coordinate of $AdS^3$ determine the natural extension 
of $SO(2,1)$ which is the isometry of $H^2$ to $SO(3,1)$ or $SO(2,2)$ so that 
the extended action of $SO(2,1)$ preserves their time-slicing, respectively. The 
identifications $\Gamma\subset SO(2,1)$ on $H^2$ also extended to 
$\gamma\subset SO(3,1),\ SO(2,2)$ on $dS^3$ and $AdS^3$, which 
preserves the time-slicing. $\gamma$ provides non-trivial topology 
$dS^3/\gamma$ or $AdS^3/\gamma$ to the 
spacetimes.

For the next section, we imbed $dS^3$ and $AdS^3$ as a  covering 
spacetime of the concerning spacetime with non-trivial topology into four 
dimensional flat spacetimes with signatures $(-+++)$ and $(--++)$, respectively.
Using the coordinate of the Poincar\'{e} model for spatial sections,
\begin{eqnarray}
	ds^2 & = & dX^2+dY^2-dZ^2
	{+\brace -}dW^2\\
	\label{eqn:imb1}
	X & = &  {\sinh t \brace \cos t}
	{2r \over 1-r^2}\cos\theta \\
	\label{}  
	Y & = & {\sinh t \brace \cos t}
	{2r \over 1-r^2}\sin\theta \\
	\label{}  
	Z & = & {\sinh t \brace \cos t}
	{1+r^2 \over 1-r^2} \\
	\label{}  
	W & = & {\cosh t \brace \sin t}
	\label{eqn:imb5}  \\
	\Longrightarrow ds^2 & = & -dt^2 + {\sinh^2 t \brace \cos^2 t} {4(dr^2+r^2 d\theta^2)\over (1-r^2)^2},
	\label{eqn:mt}
\end{eqnarray}
where the upper case is $dS^3$ and the lower case is $AdS^3$. We 
treat only the 
spacetime with a unit curvature radius for simplicity because the 
absolute value of the curvature is not 
essential for the following investigation.

\section{Quantum Field}
In this section Quantum field is investigated in the spacetime with non-trivial 
topology whose covering spacetime is $dS^3$ or $AdS^3$.
We introduce conformally coupled massless scalar field $\phi$, with the 
action
\begin{equation}
	S=-\int dx^3 \sqrt{g}\left({1\over 
	2}\partial^{\mu}\phi\partial_{\mu}\phi+{1\over 16}R\phi^2 \right),
	\label{}
\end{equation}
where $R$ is the scalar curvature.
The field equation in $dS^3$ or $AdS^3$ with $R_{\mu\nu}=\pm 2 
g_{\mu\nu}$ is
\begin{equation}
	\left(\nabla^{\mu}\nabla^{\nu}\mp{3\over 4 }\right)\phi=0
	\label{}
\end{equation}
with $R=\pm 6$ for our $dS^3$ or $AdS^3$ with a unit curvature radius. 

Now we consider the Hadamard Green functions in the covering spacetime 
$dS^3$ or $AdS^3$. According to Steif\cite{ST}, they are given by
\begin{equation}
	\bar{G}(x,y)={1\over 4\pi}{1\over \vert x-y\vert},
	\label{eqn:gc}
\end{equation}
where $\vert x-y\vert$ is a chordal distance between $x$ and $y$ in the 
four dimensional imbedding spacetime (\ref{eqn:imb1})$\sim$(\ref{eqn:imb5}) 
and not a proper distance in $dS^3$ or 
$AdS^3$.

The Hadamard functions for spacetimes with non-trivial topologies 
$dS^3/\gamma^{(i)}$, $AdS^3/\gamma^{(i)}$  can be 
obtained from the Hadamard function for their covering spacetime 
(\ref{eqn:gc})
by the method of images. Since the images of $y$ are generated by 
elements of $\gamma^{(i)}$, the Green function is
\begin{equation}
	G^{(i)}(x,y)=\sum_{T^{(i)}\in\gamma^{(i)}, T^{(i)}\neq id} \bar{G}(x,T^{(i)}\circ y).
	\label{eqn:gi}
\end{equation}
where the summation is over all elements of $\gamma^{(i)}$ except for 
identity. The identity is excluded to subtract all local contributions of 
the quantum field. This procedure ought to regularize the energy-momentum 
tensor of the quantum field.

When $\gamma$ is 
Abelian group, the summation can be easily evaluated like the three 
dimensional black hole case\cite{ST} 
(for example, $AdS^3/\gamma_BH: \gamma_BH={(\Lambda^{(1)}_1)^n}$ is equivalent to the 
three dimensional black hole). On the other hand, our non-Abelian 
$\gamma$ makes rigorous evaluations impossible.
The simple universe shown in the previous section, however, allows us to 
evaluate some divergences. The abstract summation of (\ref{eqn:gi}) is 
decomposed into   
\begin{equation}
	G^{(i)}(x,y)=\sum_{n=1}^{\infty}\sum_{j=1}^{3\cdot 4^{n-1}}\sum_k^3 
	\bar{G}(x,R(\theta_j)(\Lambda^{(i)}_k)^n\circ y)    
	\label{eqn:G}
\end{equation}
by eq.(\ref{eqn:dt})
A quantum energy-momentum tensor is given by
\begin{equation}
	<T_{\mu\nu}>=\lim_{y\rightarrow x}{\cal D}_{\mu\nu}G(x,y),
	\label{}
\end{equation}
in the point stripping method, where ${\cal D}_{\mu\nu}$ is a certain 
differential operator (see \cite{ST}, for example).
Hence, investigating the zero of the distance $\vert x-T(x)\vert$ and the summations of 
(\ref{eqn:G}), we can discuss the 
divergences of $<T_{\mu\nu}>$ since the divergences of 
$<T_{\mu\nu}>$ come out from the divergences of the Green function. 

First of all we discuss the characteristics of $\gamma^{(1)}$.
As stated in the previous section, $T^{(1)}\in\gamma^{(1)}$ can be written as
$$
T^{(1)}=A^{-1}T_h A
$$
by an appropriate $A\in SO(2,1)$, where $A$ corresponds to an isometric 
coordinate transformation of $\theta\rightarrow \theta'$ and 
$r\rightarrow r'$ in eq.(\ref{eqn:mt}). 
$T_h$ in (\ref{eqn:tx}) corresponds to a Lorentz 
boost about $X$-$Z$ direction in the imbedding spacetime. From imbedding 
equation (\ref{eqn:imb1})$\sim$(\ref{eqn:imb5}), a chordal distance 
$\vert x-T^{(1)}(x)\vert$ is given by
 \begin{eqnarray}
 	\vert x-A^{-1}T_h A\circ x\vert &=& \vert x'-T_h\circ x'\vert \\
 	&=&\sqrt{(2\cosh 2\chi-1)(X'^2-Z'^2)} \\
 	&=&{\sinh t\brace \cos t}\sqrt{(2\cosh 2\chi-1) 
 	{4r^2\cos^2\theta'-(1+r'^2)^2\over (1-r'^2)^2}},
 	\label{eqn:dt1}
 \end{eqnarray}
where $\{X',Y',Z',W'\}$ is an imbedding coordinate for $x'=A(x)$.
 $T_h$ in the imbedding spacetime $\{X,Y,Z,W\}$ is given by the matrix,
 \begin{equation}
 	\left(\begin{array}{cccc}
 		\cosh \chi & 0 & \sinh \chi & 0 \\		
 		0 & 1 & 0 & 0 \\
 		\sinh \chi & 0 & \cosh \chi & 0 \\
 		0 & 0 & 0 & 1
 	\end{array}\right).
 	\label{}
 \end{equation}
From (\ref{eqn:imb1})$\sim$(\ref{eqn:imb5}),  $\chi$ is 
given by the parameter $\beta$ of (\ref{eqn:tx}) as 
\begin{equation}
	\chi=2 \beta.
	\label{}
\end{equation}
In the $SL(2,C)$ representation (\ref{eqn:sl}), for a hyperbolic element 
$T^{(1)}$
\begin{equation}
	\chi=2 \cosh^{-1}{Tr_{SL(2,C)}(T^{(1)})\over 2}=2 
	\cosh^{-1}{a+\bar{a}\over 2},
	\label{eqn:tr}
\end{equation}
where $Tr_{SL(2,C)}$ is a trace about $SL(2,C)$ matrix. 
Especially, by the rotational symmetry of $\Lambda^{(1)}_{1\sim 3}$'s, 
all $\chi$'s of 
$\Lambda^{(1)}_{1\sim 3}$'s are the same. We write it as 
$\chi_0=2\cosh^{-1}((a_0+\bar{a_0})/2)$.

On the contrary, $T^{(2)}$ has a different standard form (\ref{eqn:t2}).
With appropriate $B\in SO(2,1)$ and $T_p$ in (\ref{eqn:t2}), $T^{(2)}$ is 
written as $B^{(-1)}T_pB$. We dare express the distance 
in the coordinate of the Poincar\'{e} model 
(\ref{eqn:imb1})$\sim$(\ref{eqn:imb5}) to evaluate it at infinity. $\vert 
x-T^{(2)}(x)\vert$ becomes
\begin{eqnarray}
\vert x-B^{-1}T_pB\circ x\vert &=& \vert x''-T_p \circ x''\vert\\
&=& (Z''+Y'')\\
&=& {\sinh t \brace \cos t} {2 r'' \sin\theta'' +1+r''^2\over 1-r''^2},
	\label{eqn:pd}
\end{eqnarray}
where $\{X'',Y'',Z'',W''\}$ and ${r'',\theta''}$ are coordinates for $x''=B(x)$.  When 
$\theta''$ is $3\pi/2$, the distance vanishes at infinity ($r''\rightarrow 
1$). This point ($r''=1, \theta''=3\pi/2$) is a fixed point of $T_p$ 
corresponding to a cusp. $T^{(2)}$ forms a cusp at $x_c$ with 
$B(x_c)=(r''=1,\theta''=3\pi/2)$. 
Then $\vert x-T^{(2)}(x)\vert$ should vanish at the cusp on 
each timeslice, while $\vert x-T^{(1)}(x)\vert$ never approaches 
to zero except for the initial or final singularity.

There are three possibilities of divergences for the quantum field. The quantum field will diverge at 
the singularity of the background spacetime as in the case of three 
dimensional black hole\cite{ST}. Also the infinitely small structure of 
cusps will cause the divergence of the quantum field though the cusp is 
not a singularity. Furthermore, the summation of the images in the 
method of images is expected to diverge since the volume of the 
hyperbolic space is larger than that of the flat space at a distant region. 
One will see so many images of a source.

First two possibilities are investigated in the following. 
 For $T^{(1)}\in \gamma^{(1)}$, eq.(\ref{eqn:dt1}) tells that 
 $\vert x-T^{(1)}(x)\vert$ vanishes at the planes $X'^2-Z'^2=0$. 
Since $A\in SO(2,1)$ is a proper-Lorentz group 
in a part of the imbedding spacetime $\{X,Y,Z,W=\pm1\}$, the planes $X'^2-Z'^2=0$ cannot 
invade into the inside of a light cone at $\{0,0,0, \pm1\}$ which is 
initial or final singularity of $dS^3$ and $AdS^3$. Moreover, 
one or two pairs of null geodesic, $\{X'^2-Z'^2=0, Y'=0, W'= \pm1\}$, lie on 
the singularities. Since the quantum field diverges on these planes  
$<T_{\mu\nu}>$ diverge when one approaches to the singularity of the 
background spacetimes by the 
topological effect.

On the other hand, $T^{(2)}\in \gamma^{(2)}$ has a different 
characteristic about cusps. 
From eq.(\ref{eqn:pd}), $\vert x-T^{(2)}(x)\vert$ vanishes at the cusps on each 
time-slice, while $\vert x-T^{(1)}(x)\vert$ never vanishes on each 
time-slice even at their cusps. Therefore $<T_{\mu\nu}>$ is singular at 
the cusps of $dS^3/\gamma^{(2)}$ and $AdS^3/\gamma^{(2)}$ and regular at 
the cusps of $dS^3/\gamma^{(1)}$ and $AdS^3/\gamma^{(1)}$.

Finally we discuss the third possibility of the divergences estimating 
the summation  over all transformations in (\ref{eqn:G}). From a 
rotational symmetry around the origin of each time-slice $(X=Y=0,Z={\sinh 
t\brace \cos t},W={\cosh t\brace \sin t})$,
\begin{eqnarray}
	\vert x-R(\theta)(\Lambda^{(1)}_k)^n\circ x\vert_{x=origin}&=&	\vert 
	x-(\Lambda^{(1)}_k)^n\circ x\vert_{x=origin} \\
	&=&{\sinh t \brace \cos t}\sqrt{2\cosh(2n\chi_0)-1},
	\label{}
\end{eqnarray}
where we use $A^{-1}\Lambda^n A=(A^{-1}\Lambda A)^n$. If we consider the 
point $x=x_0$,
\begin{eqnarray}
	{\sinh t \brace \cos t}\sqrt{2\cosh(2n\chi_0)-1}-2\vert x_0\vert&\leq&\vert 
	x-R(\theta)(\Lambda^{(1)}_k)^n\vert_{x=x_0}\\
	&\leq&{\sinh t \brace \cos t}\sqrt{2\cosh(2n\chi_0)-1}
	+2\vert x_0\vert
	\label{}
\end{eqnarray}
For a sufficiently large $n$, $\sqrt{2\cosh(2n\chi_0)-1}$ and $\vert 
x-R(\theta)(\Lambda^{(1)}_k)^n(x)\vert$ behaves as $e^{n|\chi_0|}$. Furthermore, $\vert 
x-R(\theta)(\Lambda^{(2)}_k)^n\vert$ also behaves as $e^{n|\chi_0|}$ for a 
large $n$, since the 
tessellation of $\gamma^{(2)}$ is the same tessellation as $\gamma^{(1)}$.
From (\ref{eqn:n1})(\ref{eqn:n2})(\ref{eqn:n3})(\ref{eqn:tr}) the summation about $j$ and $k$ in (\ref{eqn:G}) is estimated as
\begin{eqnarray}
\lim_{y\rightarrow x}\sum_{j=1}^{3\cdot 4^{n-1}}\sum_k^3 
\bar{G}(x,R(\theta_j)(T_k)^i\circ y)\sim {4^n\over e^{n|\chi_0|}}=\left({4\over 4.74}\right)^n=0.8438^n.   
	\label{}
\end{eqnarray}
Though a rigorous estimation may be possible, it is too complicated and 
will give us no essential information.
$<T_{\mu\nu}>\sim O(\sum_n 0.843^n)$ barely converge because of the exact 
value of $\chi_0$. If the investigation could be done in other topology 
with negative curvature, a different value of $\chi$ might cause the 
divergences of the summation.
\section{Summary and Discussion}
In the present article, we have investigated new topological effects of a quantum field 
in a torus universe with a cusp and a sphere universe with three cusps. 
Their covering spacetime is de Sitter spacetime or anti de Sitter 
spacetime. The cusp is a point at infinity with regular local structure 
and needle-like global structure.
Three possibilities of divergences of the energy-momentum tensor have been studied. 

First, a divergence
appears on the coordinate singularity of the classical background spacetime, which 
is initial or final singularity in cosmological sense.  This is similar to the case of the three dimensional black hole\cite{ST}.

Next possibility is a divergence at the cusps. In the present article, we 
show there are two 
types of the cusps. One is a cusp made by hyperbolic transformations of 
$SO(2,1)$ and the other is made by a parabolic transformation. We observed 
that $<T_{\mu\nu}>$ diverge at the latter cusps, which are included in the 
sphere with three cusps. This aspect means that the latter cusp is quantum mechanically unstable. Only the latter cusps will require a treatment in quantum gravity.

The last possibility is the divergence of the summation of images. This corresponds 
to the effect that we see more images of a source in a negatively curved universe 
than in a flat universe at a distant region. The summation, however, 
converges in the spacetimes given in the present article. The convergence of the image 
summation strongly depends on the values of a 
boost angle $|\chi|$ and the shape of a tessellation in the covering space. Though $e^{|\chi_0|}$ is 4.74, if $e^{|\chi_0|}$ were less 
than 4 with the same tessellation, $<T_{\mu\nu}>$ would diverge everywhere and the divergence is 
hard to remove. In the case of other topology, other 
$e^{|\chi|}$ and other tessellation may make the summation diverge. If so, it will turn out that there 
are topologies accepting quantum field and not accepting. When we 
consider a compact (without a cusp) topology with a negative curvature, 
such situation may occur, though the compact topology is very difficult to treat.

Recently Brill\cite{BR} shows that a three dimensional multi-black hole solution 
can be constructed in the three dimensional anti-de Sitter spacetime. It 
is easily found that $AdS^3/\gamma^{(2)}$ with a larger boost angle 
$|\chi'|$  than $|\chi_0|$ is regarded as a two-black hole solution. Of 
course, the summation of images converges in this solution.

By a regularization 
performed in the present article, we perfectly subtract local divergences. We, 
however, have observed the topological divergences. They cannot be 
regularized and will have physical meanings.

Can we carry out a similar investigation in other 
topology with a negative curvature. A compact topology (without a cusp) seems impossible since the 
tessellation is so complicated. On the other hand, it may be possible to 
treat other topologies with cusps. At least, we can decide whether each 
cusp causes a divergence of quantum field or not knowing whether the 
identification providing the cusp is parabolic or hyperbolic. The divergence of summation of images 
sensitively depends on the shape of the tessellation in the covering 
space and will be 
difficult to treat without a sufficient symmetry. In (3+1)-dimension 
the similar investigation may be possible. There will be convenient 
models of non-trivial topology.

\acknowledgements
We would like to thank Professor H. Sato and Dr. T. Tanaka for helpful discussions.
The author thanks the Japan Society for the 
Promotion of Science for financial support. This work was supported in 
part 
by the Japanese Grant-in-Aid for Scientific Research Fund of the Ministry 
of 
Education, Science, Culture and Sports.


\end{document}